\documentclass[aps,prd,superscriptaddress,twocolumn,10pt,amsmath,amssymb,nofootinbib]{revtex4-1}
\usepackage{acronym}
\usepackage{bm}
\usepackage[dvipdfmx]{graphicx}
\usepackage{subfigure}
\usepackage[usenames]{color}
\usepackage{url}

\definecolor{mygreen}{RGB}{0,130,0}

\newcommand{\beq}{\begin{equation}}
\newcommand{\eeq}{\end{equation}}
\newcommand{\bea}{\begin{eqnarray}}
\newcommand{\eea}{\end{eqnarray}}

\usepackage{hyperref}

\newcommand*{\diff}{\,\mathrm{d}}
\newcommand{\abs}[1]{\left\lvert #1 \right\rvert}
\newcommand{\av}[1]{\left\langle #1 \right\rangle}
\newcommand{\Omegagw}{\Omega_{\textrm{GW}}}

\newcommand{\figref}[1]{Figure~\ref{#1}}
\newcommand{\tabref}[1]{Table~\ref{#1}}
\newcommand{\secref}[1]{Sec.~\ref{#1}}

\begin{document}


\title{Subtracting compact binary foregrounds utilizing anisotropic statistic for third-generation gravitational-wave detectors\\
}

\author{Soichiro Kuwahara}
\affiliation{Research Center for the Early Universe (RESCEU), Graduate School of Science, The University of Tokyo, Tokyo 113-0033, Japan}
\author{Atsushi Nishizawa}
\email{Corresponding author: atnishi@hiroshima-u.ac.jp}
\affiliation{Physics Program, Graduate School of Advanced Science and Engineering, Hiroshima University, Higashi-Hiroshima, Hiroshima 739-8526, Japan}
\affiliation{Astrophysical Science Center, Hiroshima University, Higashi-Hiroshima, Hiroshima 739-8526, Japan}
\author{Lorenzo Valbusa Dall'Armi}
\affiliation{Dipartimento di Fisica ``Enrico Fermi'', Universit\`a di Pisa, Largo Bruno Pontecorvo 3, Pisa I-56127, Italy}
\affiliation{INFN, Sezione di Pisa, Largo Bruno Pontecorvo 3, Pisa I-56127, Italy}

\begin{abstract}
The astrophysical foreground from compact-binary coalescence signals is expected to be a dominant part of total gravitational wave (GW) energy density in the frequency band of the third-generation detectors. The detection of any other subdominant stochastic GW background (GWB), especially a primordial GWB, will be disturbed by the astrophysical foreground, which needs to be cleaned for further studies of other stochastic GWB. Although previous studies have proposed several cleaning methods, the foreground from subthreshold binary neutron stars (BNS) has been a major obstacle to remove. In this paper, we propose the novel idea to acquire better estimation of the unresolved foreground, by using the information about its anisotropies. We simulate the BNS population and compute its angular power spectrum and shot noise. We find that the shot noise from BNS is too faint to observe after subtracting loud signals due to the limited angular resolution of the third-generation detectors. This justifies the approximation regarding the unresolved foreground as an isotropic component. We also discuss the angular resolution necessary to make our method valid for the foreground subtraction.
\end{abstract}

\maketitle


\section{Introduction}
After the first direct observation of gravitational wave (GW) from the coalescence of binary black holes in 2015 \cite{GW150914}, ground-based detectors have observed many compact-binary coalescence (CBC) signals. The LIGO, Virgo, and KAGRA collaboration published the catalog of CBC candidates along with the observation period such as GWTC-2 \cite{GWTC-2} and GWTC-3 \cite{GWTC-3}. These observations imply that there is a population system of CBC and one can expect there are more CBC with weak amplitude such that they are not individually detected. The superposition of these CBC would compose a stochastic GWB \cite{Regimbau2008,Rosado2011, Wu2012,Zhu2013,Zhu_2011,Marassi_2011}. The other astrophysical sources contributing to the stochastic GWB are, for example, magnetars \cite{Regimbau_2001,Marassi_2010,Wu_2013} and core-collapse supernovae \cite{Buonanno2005,Howell_2004,Sandick_2006,Marassi_2009,Zhu_2010}. The other potential sources of stochastic GWB are cosmological sources. One can point out cosmic strings \cite{Sarangi_2002,Damour_2005,Siemens_2007,Kibble_1976}, GW emitted during the inflationary era \cite{Starobinsky_1979,Turner_1997,Bar_1994}, and primordial black holes \cite{Sasaki_2016,Mandic_2016,Wang_2018}.

Current second-generation ground-based interferometric detectors such as LIGO \cite{DetectorALIGO}, Virgo \cite{DetectorAVirgo}, and KAGRA \cite{DetectorKAGRA} have not detected any stochastic GWB signals \cite{O1Isotropic,O2Isotropic,O3Isotropic}. However, increased sensitivity in the third-generation (3G) ground-based detectors including Cosmic Explorer (CE) \cite{DetectorCE} and Einstein Telescope (ET) \cite{DetectorET1,DetectorET2} might achieve its detection. The stochastic GWB in the frequency window of these detectors is expected to be dominated by CBC origin. The CBC foreground would hinder the investigation of any other subdominant stochastic GWBs \cite{Regimbau_2017,Belgacem_2024}. Therefore, for the sake of probing subdominant stochastic GWB, techniques of foreground subtraction have been discussed. 

There are several studies on techniques and their consequences of foreground subtraction. The main inhibiting factors for this subtraction process are (1) residual errors and (2) unresolved signals. 

(1) An imperfect fit of the waveform causes residual remnants after subtraction. The latest studies \cite{Zhou_2022,Zhou_2023,Song_2024} have indicated that residual errors caused by fitting the waveform with a higher dimension of the parameter space are greater than predicted in \cite{Sachdev_2020} and greatly affect the subtraction. Several studies \cite{Cutler_2006,Sharma_2020,Pan_2023} introduce a projection method to reduce the level of residual errors that applies a projection operator to residual data and removes components that lie within the tangent space of the manifold at the point corresponding to the best-fitting waveform. As an alternative way, Ref.\@\cite{Zhong_2023,Zhong_2024} have proposed notching procedures on time-frequency spectrogram for foreground subtraction. They notch out pixels containing CBC signals by the mask. Although residual errors on this method are sufficiently small in their method from 10 to 100\,Hz, they leave most of the pixels at lower frequencies below 10\,Hz. All these recent studies are carried out to reduce residual errors, and unresolved signals still remain a major obstacle in searching stochastic GWB other than CBC background. 

(2) Weak signals below the threshold would not be subtracted, creating an unresolved background. The level of unresolved background by binary neutron stars (BNS) is louder than the one by binary black holes with three or five 3G detectors network \cite{Sachdev_2020,Zhou_2022,Zhou_2023}. This is due to the fact that the masses of BNS are smaller than those of binary black holes and their signal-to-noise ratio (SNR) are lower. To deal with the foreground by unresolved binaries, some studies have tried to estimate the unresolved foreground from resolved signals \cite{Li_2025,Zhong:2025qno}. These searches have assumed that the unresolved CBC foreground is isotropic, which is not true in reality.

In this paper, we focus on the BNS foreground for the difficulty of unresolved binaries mentioned above, and propose the novel method for searching isotropic cosmological GWB under the subtraction process in the frequency domain using information from the anisotropic stochastic GWB. The finite sampling of CBC in limited observation time would induce the shot noise to anisotropy and the angular power of the shot noise would be intimately connected with the number of signals and the power distribution of the signals. Therefore, once we could estimate the local merger rate and the redshift and mass distribution of the detected events, we could formalize the relationship between the angular power of the CBC background and isotropic energy density of CBC background, meaning that the observation of anisotropy can be used to better estimate the isotropic component of the unresolved foreground. As discussed in \cite{Zhong_2024,Zhong:2025qno}, subthreshold BNS plays a significant role in the strength of the unresolved foreground. Therefore, we simulate BNS population and investigate the sensitivity to the isotropic cosmological GWB using information from the anisotropic stochastic GWB.

We begin with the explanation of our astrophysical population model and the computation of the angular power spectrum originating from the shot noise in the source number in \secref{sec:catalog}. We derive our likelihood function and define the parameters in \secref{sec:likelihood_estimator}. Especially, we define how we describe the improvement in the estimation of the energy density of the cosmological GWB in \secref{sec:new_estimator}. In \secref{sec:results}, we demonstrate how the angular power spectrum of the shot noise is altered by the astrophysical population models and show the result of parameter estimation with three detector networks of the 3G detectors. We further discuss future prospects of our method with higher angular resolution detectors in \secref{sec:Discussion}.

\section{Astrophysical population and mock catalog}
\label{sec:catalog}
In this section, we introduce the astrophysical population model for BNS, which we use in the analysis. For the cosmological parameters, we apply the values of those determined by Planck 2018 \cite{Planck18}.

\subsection{Redshift distribution}
\label{sec:redshift_dist}
We simulate an astrophysical population up to redshift $z=10$ because redshifts larger than $10$ contribute little to the total background \cite{GW170817_SGWB}. For the computation of the redshift-dependent merger rate, we follow the same procedure as described in \cite{Regimbau_2017,Sachdev_2020,Zhou_2023}. The redshift dependence of the merger rate is given by the merger rate in the observer frame $R(z)$:
\begin{align}
    \label{eq:redshift_dist}
    p(z) = \frac{R(z)}{\int_0^{z_{\rm max}} R(z)\diff z} \;,
\end{align}
where we set $z_{\rm max}=10$. The merger rate in the observer frame is
\begin{align}
    R(z) = \frac{R_{\rm m}(z)}{1+z}\frac{\diff V}{\diff z}(z)
    \label{eq:merger_rate_observer_frame}
\end{align}
where $\diff V/\diff z$ is the comoving volume element and $R_{\rm m}(z)$ is calculated as follows.

We assume that the binary formation rate is proportional to the star formation rate per observer's time per comoving volume, given in \cite{bestfitSFR_Vangioni2015}:
\begin{align}
    R_{\rm f}(z) = \nu\frac{ae^{b(z-z_m)}}{a-b+be^{a(z-z_m)}}
\end{align}
where $\nu=0.146\,M_{\odot}\,{\rm Gpc}^{-3}\,{\rm yr}^{-1}$, $z_m=1.72$, $a=2.80$, and $b=2.46$. The functional form is called the Springer-Hernquist functional form \cite{Hernquist_Springel} and the parameters are obtained by fitting the data based on the gamma-ray burst in \cite{kistler2013fromGRB}. The normalization process from the gamma-ray burst rate to the star-formation rate is based on \cite{Trenti2013norm,Behroozi2015norm}. We also assume that the probability distribution of the time delay $t_{\rm d}$ between the formation of a binary and its merger follows $p(t_{\rm d}) \propto 1/t_{\rm d}$ in the range of $(t_{\rm d}^{\mathrm{min}},t_{\rm d}^{\mathrm{max}})$. We set $t_{\rm d}^{\mathrm{min}}=20\,{\rm Myr}$ \cite{GW170817_SGWB,Meacher_2015} and $t_{\rm d}^{\mathrm{max}}$ to be the Hubble time. We can obtain the merger rate per comoving volume in the source frame from these assumption as follows,
\begin{align}
    R_{\rm m}(z) = \int_{t_{\rm d}^{\mathrm{min}}}^{t_{\rm d}^{\mathrm{max}}} R_{\rm f}(z_{\rm f})p(t_{\rm d})\diff t_{\rm d}, \quad z_{\rm f} = z[t(z)-t_{\rm d}]
    \label{eq:merger_rate_source_frame}
\end{align}
where $z$ is the redshift of the merger, $t(z)$ is the cosmic time of the merger, and $z_{\rm f}$ is the redshift at the binary formation time calculated from $t(z)-t_{\rm d}$. The $R_{\rm m}(z)$ is calibrated by the local merger rate $R_{\rm m}(z=0)$, determined from the observations.

\subsection{Local rate and mass distribution}
We assume a simple uniform mass distribution between $1\,M_{\odot}$ and $2.5\,M_{\odot}$ as a fiducial model, which was also chosen as a fiducial model in the previous search \cite{MR_GWTC2} with the GWTC-2 catalog \cite{GWTC-2}. They infer the local merger rate of $320^{+490}_{-240}\,{\rm Gpc}^{-3}\,{\rm yr}^{-1}$ assuming this mass distribution.
We also performed the analysis with different mass distributions, a Gaussian distribution with mean $1.33\,M_{\odot}$ and standard deviation $0.09\,M_{\odot}$, adopting the one introduced in \cite{Sachdev_2020} to examine the effect of a different mass distribution.
All population models that we applied for this analysis are shown in \tabref{tab:popularion_pattern}.
\begin{table}[]
    \centering
    \begin{tabular}{|l||c|c|} \hline
      & local rate & mass distribution \\ \hline
     pop Nmed & $320\,{\rm Gpc}^{-3}\,{\rm yr}^{-1}$ & $\mathcal{N}(1.33\,M_{\odot},\,8.1e-4\,M_{\odot}^2)$ \\ \hline
     pop Umed & $320\,{\rm Gpc}^{-3}\,{\rm yr}^{-1}$ & $\mathcal{U}(1.0\,M_{\odot},\,2.5\,M_{\odot})$ \\ \hline
     pop Ulow & $80\,{\rm Gpc}^{-3}\,{\rm yr}^{-1}$ & $\mathcal{U}(1.0\,M_{\odot},\,2.5\,M_{\odot})$ \\ \hline
     pop Uhigh & $810\,{\rm Gpc}^{-3}\,{\rm yr}^{-1}$ & $\mathcal{U}(1.0\,M_{\odot},\,2.5\,M_{\odot})$ \\ \hline
    \end{tabular}
    \caption{Population models used in this paper. $\mathcal{N}(\mu,\sigma^2)$ represent a Gaussian distribution with mean $\mu$ and standard deviation $\sigma$ and $\mathcal{U}(a,b)$ represents a uniform distribution.}
    \label{tab:popularion_pattern}
\end{table}

\subsection{Mock catalog generation}

\subsubsection{Number of events, redshift, and sky location}
We computed the expected number of mergers in one year of observation in a certain redshift window $[z-\frac{1}{2}\diff \ln z, z+\frac{1}{2} \diff \ln z]$ from Eq.~\eqref{eq:merger_rate_observer_frame}. We created 100 bins uniformly in a logarithmic space ranging from $z_{\rm min}=0.05$ to $z_{\rm max}=10$. Then we compute the expectation value $\langle n_i \rangle$ for each bin. The number of mergers in each bin is drawn randomly from the Poisson distribution with the mean $\langle n_i \rangle$. A same redshift is assigned for binaries in a single bin.

The position in the sky is given by the azimuthal angle $\phi$ and the polar angle $\theta$. We let the distribution $\phi$ be a uniform distribution $[0,2\pi]$ and $\cos\theta$ be distributed uniformly in $[-1,1]$.

\subsubsection{Waveform and SNR computation}
\label{sec:SNR computation}

We assume the network of three third generation detectors consisted of ET and CE and set the locations of the 3G detectors as follows: ET \cite{DetectorET1,DetectorET2} at the Virgo site\footnote{The ET location is considered to be either of Sardinia in Italy and the Euregio Meuse-Rhine at the Netherlands-Belgium border~\cite{DiGiovanni:2025oyo}. However, it has not been decided and we put it at the Virgo site for this study.} , CE with $20\,{\rm km}$ arms \cite{DetectorCE} at the LIGO Hanford site, and CE with $40\,{\rm km}$ \cite{DetectorCE} arm at the LIGO Livingston site. 
They have good sensitivity in the range of several tens to several hundreds of hertz, we choose the reference frequency at $24.77\,{\rm Hz}$ and consider the frequency range for the analysis from $10\,{\rm Hz}$ to $200\,{\rm Hz}$. The frequency band of interest is entirely dominated by the inspiral phase of CBC signals in the case of BNS and we here use inspiral waveform. Taking into account the antenna response, we use the angular-averaged waveform in the frequency domain \cite{Cutler_1994,Poisson_1995}
\begin{align}
    \tilde{h}(f) = \sqrt{\frac{1}{30}}\frac{1}{\pi^{2/3}}e^{i\psi_{+}(f)}\frac{c}{d_{\rm L}(z)}\left(\frac{G\mathcal{M}_{z}}{c^3}\right)^{5/6}f^{-7/6},
    \label{eq:inspiral_response}
\end{align}
where $\mathcal{M}_z\equiv(1+z)\mathcal{M}$ is the redshifted chip mass, $d_{\rm L}(z)$ is the luminosity distance, and $\psi_{+}(f)$ is the phase, which is not important for the computation of SNR.

Using Eq.~\eqref{eq:inspiral_response}, we can compute SNR of each binary by the formula
\begin{align}
      \left(\frac{S}{N}\right)^2  &= 4\int_{f_{\rm min}}^{f_{\rm max}}\diff f\frac{|\tilde{h}(f)|^2}{S_n(f)}.
\end{align}
The $f_{\rm min}$ is set to the lowest frequency at which we can obtain the ET and CE's power spectral density from publicly available data \cite{detector_noise_curve_ET_CE}. To put it specifically, $f_{\rm min}=1\,{\rm Hz}$ for ET and $f_{\rm min}=5\,{\rm Hz}$ for CE. The The $f_{\rm max}$ is the frequency of the innermost stable circular orbit (ISCO) frequency of the Schwarzschild black hole, $f_{\rm ISCO} = (6^{3/2}\pi M_z)^{-1}$ with the redshifted total mass $M_z \equiv (1+z)(m_1+m_2)$. Since we assume the detector network of three 3G detectors, we compute network SNR as the root sum square of SNR for individual detector $I$, 
\begin{align}
    \mathrm{SNR}_{\rm net} = \sqrt{\sum_I {\rm SNR}_{I}^2}\;.
\end{align}

\begin{figure}
    \centering    \includegraphics[width=\linewidth]{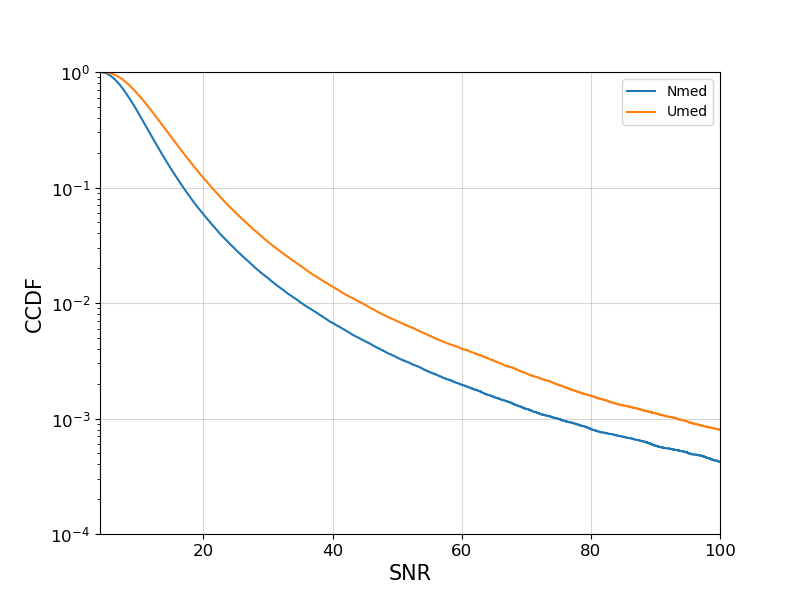}
    \caption{The complementary cumulative distribution function (CCDF) for SNR catalog of BNS. This CCDF is created from one realization of a BNS population with the model pop Nmed and Umed. It indicates how much fraction of events when we count the events from the highest to the given SNR.}
    \label{fig:CCDF-SNR}
\end{figure}

\begin{figure}
    \centering
    \includegraphics[width=\linewidth]{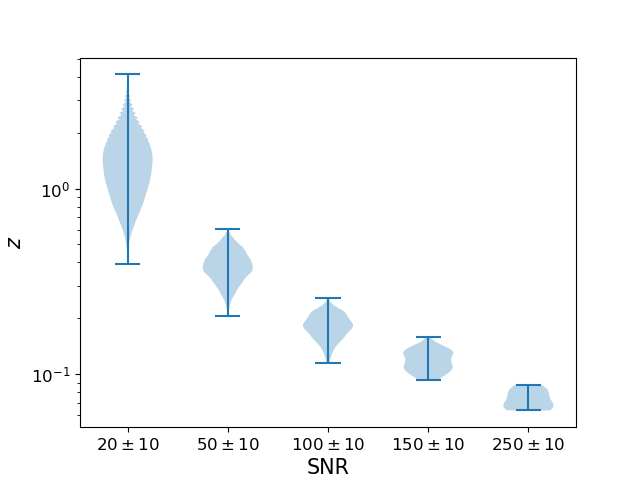}
    \caption{The violin plot of redshift distribution in each SNR bin. We used one out of 100 realizations of the pop Umed model.}
    \label{fig:snr_z_violin}
\end{figure}

 The resultant SNR distribution is given in \figref{fig:CCDF-SNR}. In \figref{fig:snr_z_violin}, the redshift distribution in each SNR bin is shown as a violin plot. This indicates that fewer loud signals located at low redshifts contribute mainly to the anisotropy.
 
\subsection{Gravitational-wave background spectrum}

The value of $\Omegagw$ from a population of CBCs can be written as \cite{Regimbau_2017,Zhong_2023,Sachdev_2020},
\begin{align}
    \label{eq:flux_and_omegagw}
    \Omega_{\rm GW}(f) = \frac{1}{\rho_c c}fF(f)\;,
\end{align}
where $F(f)$ is the total energy flux as sum of individual sources. The total energy flux is
\begin{align}
    \label{eq:total_flux_CBC}
    F(f) = \frac{1}{T}\frac{\pi c^3}{2G}f^2\sum_{i=1}^N\left[\abs{\tilde{h}^{(i)}_{+}(f)}^2+\abs{\tilde{h}^{(i)}_{\times}(f)}^2\right] \;,
\end{align}
where the index $i$ is associated with each individual source and the summation is taken over all $N$ sources in the observation time $T$.

\begin{figure}
    \centering
\includegraphics[width=\linewidth]{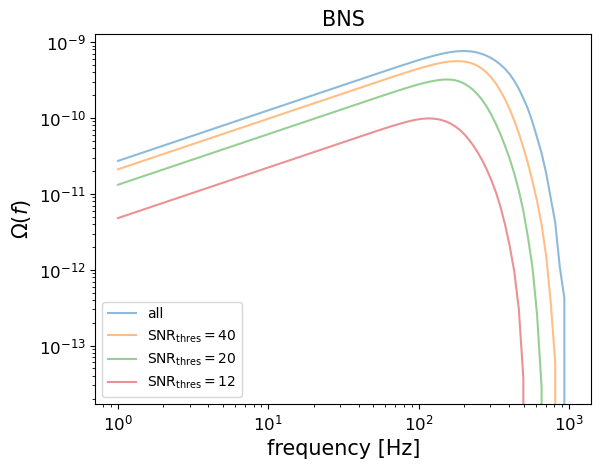}
    \caption[The energy density spectrum of unresolved CBC background for pop Umed model.]{The energy density spectrum of CBC background for pop Umed model. The blue line shows the total energy density spectrum while other yellow, green, and red line represent the energy density of unresolved background with given SNR threshold.}
    \label{fig:Omegagw_popUmed_SNRth}
\end{figure}

\begin{figure}
    \centering  \includegraphics[width=\linewidth]{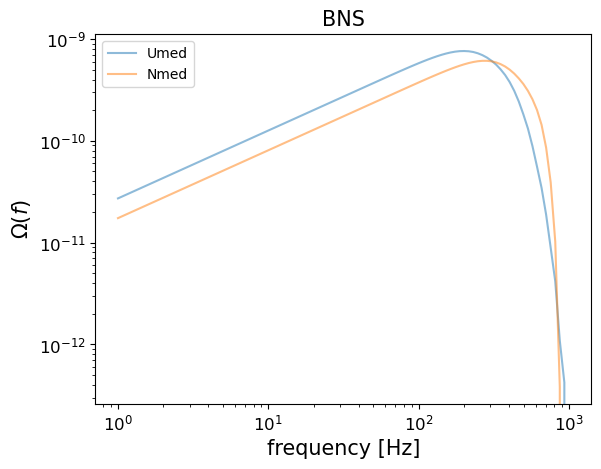}
    \caption[The energy density spectrum of unresolved CBC background for pop Umed and Nmed models.]{The energy density spectrum of CBC background for pop Umed model and Nmed models. }
    \label{fig:Omegagw_popUmed_popNmed}
\end{figure}

\figref{fig:Omegagw_popUmed_SNRth} is an example of the resultant energy density spectrum of a CBC background. We show the computational result of one realization of the signals in a year of observation with pop Umed model. The figure shows that as we lower the SNR threshold that delineates unresolved and resolved binaries, the total energy density of the unresolved binaries decreases. Since we only use the inspiral waveform up to $f_{\rm ISCO}$ for our simulation, there are steep decreases in the energy density over a few hundred $\,Hz$. However, since our choice of reference frequency is about $25\,{\rm Hz}$, $f_{\rm ISCO}$ does not affect the total energy density at the reference frequency at which the spectrum follows the power law of the index $2/3$. In Fig.~\ref{fig:Omegagw_popUmed_popNmed}, we show the difference of the energy density spectra for the pop Umed and Nmed models.

\subsection{Anisotropy of gravitational-wave backgrounds}

We assume a stationary, unopolarized, Gaussian, and stochastic GWB. With these assumptions, we define an anisotropic power spectral density of a GW background, $\mathcal{P}(f,\bm{\hat{n}})$, by the expectation value of quadratic GW strain $h_{A}(f,\bm{\hat{n}})$ at frequency $f$, directing in the unit vector of $\bm{\hat{n}}$ and having a polarization $A$,
\begin{align}
    \av{h_{A}^*(f,\bm{\hat{n}})h_{A'}(f',\bm{\hat{n}}')} = \frac{1}{4}\mathcal{P}(f,\bm{\hat{n}})\delta(f-f')\delta_{AA'}\delta^2(\bm{\hat{n}},\bm{\hat{n}}')\;,
    \label{eq:quadratic_av_anisotropic}
\end{align}
where the asterisk denotes the complex conjugate. We generally assume $\mathcal{P}(f,\bm{\hat{n}})$ as the product of two functions of spectral and spatial components:
\begin{align}
    \mathcal{P}(f,\bm{\hat{n}}) = \bar{H}(f)\mathcal{P}(\bm{\hat{n}})\;,
\end{align}
where the function $\bar{H}(f)$ is a dimensionless quantity normalized at the reference frequency as 
\beq
\bar{H}(f)= \left( \frac{f}{f_{\rm ref}} \right)^{\alpha-3} \;.
\eeq
We applied $f_{\rm ref}=24.77\,{\rm Hz}$ in this analysis.

The direction-dependent power spectrum can be decomposed into the spherical harmonic components as
\beq
\mathcal{P}(\bm{\hat{n}}) = \sum_{l=0}^\infty \sum_{m=-l}^l \mathcal{P}_{l m} Y_{l m} (\bm{\hat{n}}) \;,
\eeq
with each harmonic coefficient
\beq
\mathcal{P}_{l m} = \int \diff \bm{\hat{n}}\, \mathcal{P}(\bm{\hat{n}}) Y_{l m}^* (\bm{\hat{n}}) \;.
\eeq
The angular power spectrum is defined by
\beq
\langle \mathcal{P}_{l m}^* \mathcal{P}_{l^{\prime} m^{\prime}} \rangle \equiv C_l\, \delta_{l l^{\prime}} \delta_{m m^{\prime}} \;.
\eeq
The isotropic component, $C_0$, is related to the GW energy density by~\cite{Thrane_2009},
\beq
\Omega_{\mathrm {GW}}(f_{\rm ref}) = \frac{2\pi^2}{3H_0^2} {f_{\rm ref}^3} \sqrt{4\pi C_0}\label{eq:OmegaGW-C0}
\eeq
and we can connect to $\Omega_{\rm GW}(f)$ by $\Omega_{\rm GW}(f)=\Omega_{\rm GW}(f_{\rm ref})\left(f/{f_{\rm ref}}\right)^{\alpha}$.

To obtain angular power spectrum $C_l$, we randomly generate sources in the sky and compute the values $C_l/C_0$ and $C_0$ separately. This is because our purpose to generate $C_l$ is not investigating $C_l$ itself but extracting the information about an isotropic component from $C_l$. We computed the $C_l$ and $\delta C_l$ up to $l=6$ from 100 realizations of the distribution of the merger events in the sky in a year of observation.

\begin{figure*}    
\includegraphics[width=.48\linewidth]{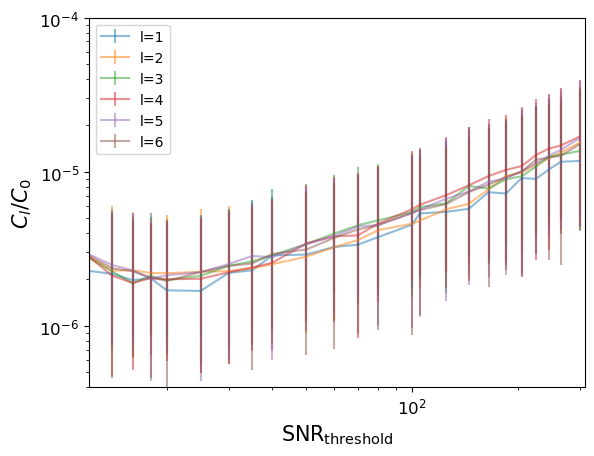}
\includegraphics[width=.48\linewidth]{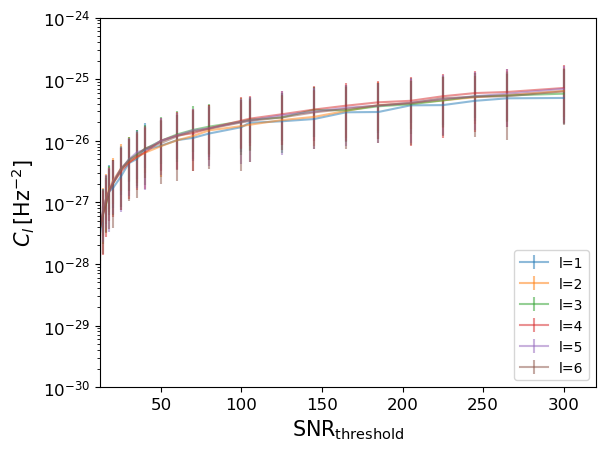}
\caption{Left: the ratio of $C_l$ to $C_0$ in a mode $l$ as a function of SNR threshold. Right: angular power spectra $C_l$ at the reference frequency $24.77\,{\rm Hz}$ as a function of SNR threshold for the population model, the pop Umed. We assume the signals whose SNR are above the threshold are subtracted perfectly from the data and only the weaker signals are counted for an anisotropic GWB.}
\label{fig:Cl320u}
\end{figure*}

\figref{fig:Cl320u} are the plots of the ratio of $C_l$ to $C_0$ (left) and the value of $C_l$ (right) as functions of the SNR threshold for the pop Umed model. We assume that signals louder than the thresholds are perfectly subtracted and keep the power of only subthreshold signals in $C_l$ and $C_0$. The error bars are from the 25th percentile to the 75th percentile. This variation in $C_l$ values is called cosmic variance and we can access to only one realization through an observation. In the left panel of \figref{fig:Cl320u}, $C_l/C_0$ is minimum at a certain SNR threshold as listed in Table~\ref{tab:ClminSNR} and increases monotonically as the SNR threshold is larger. This indicates that louder signals contribute more to anisotropy. On the other hand, when the SNR threshold is sufficiently low that most of the BNS signals are subtracted, the number of signals remaining in the data is too small and the anisotropic contribution relative to the isotropic one increases. In the right panel of \figref{fig:Cl320u}, the amplitude of $C_l$ keeps decreasing as the SNR threshold decreases. This is simply because more signals are subtracted and the remaining total energy reduces.

\begin{figure*}
    \centering
\includegraphics[width=.48\linewidth]{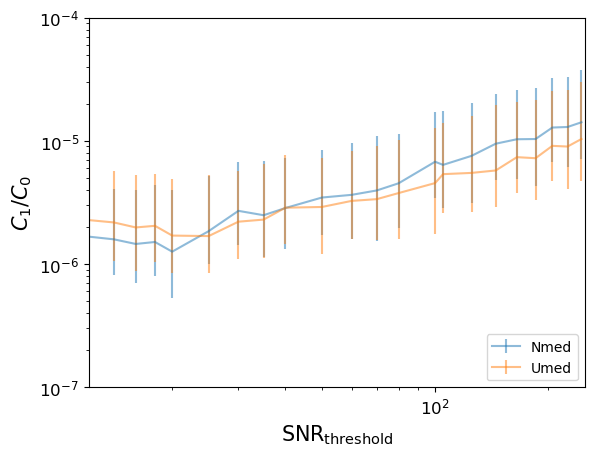}
\includegraphics[width=.48\linewidth]{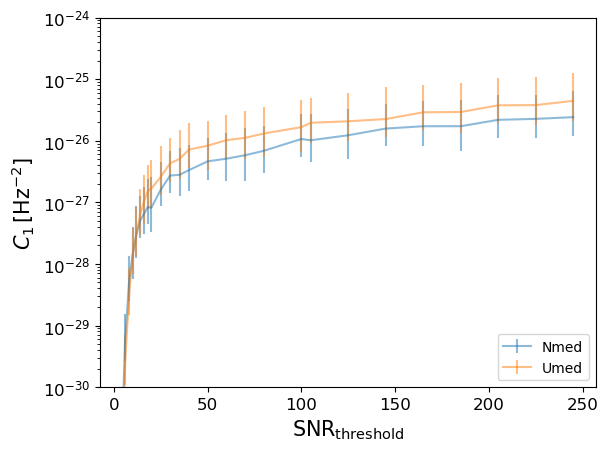}
    \caption{Same as Fig.~\ref{fig:Cl320u} but for different mass distributions. We apply the local rate of $320\,{\rm Gpc}^{-3}\,{\rm yr}^{-1}$ while mass distribution is either uniform distribution or Gaussian distribution described in \tabref{tab:popularion_pattern}. Although we calculated 100 realizations and compute angular power up to $l=6$, here we only show the median value of the $l=1$ case out of 100 realizations.}
    \label{fig:Cl_mass}
\end{figure*}

The results above are robust against the change of the population model from the uniform one to the Gaussian one. In \figref{fig:Cl_mass}, $C_l/C_0$ and $C_l$ are plotted for the models with the same local rate but the different mass distributions. The pop Umed model can generate relatively louder signals than in the pop Nmed model. It causes larger angular power because the total energy density spectrum $\Omegagw$ is larger. The plot on the left shows that $C_l/C_0$ for the pop Nmed model is slightly larger, but this is simply because $C_0$ is larger for the pop Umed model. The difference is within the error bars and is insignificant. 

\begin{figure*}
    \centering
\includegraphics[width=.48\linewidth]{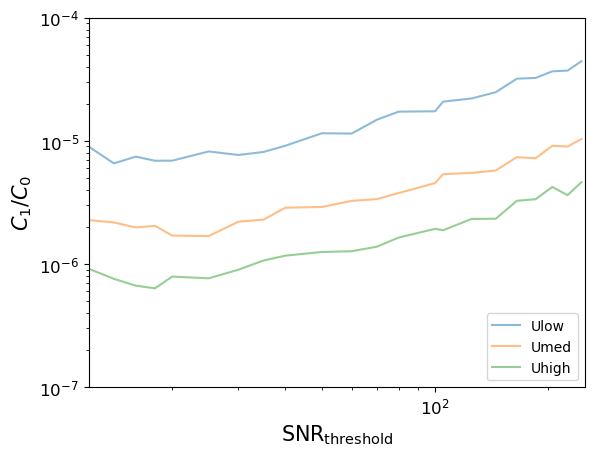}
\includegraphics[width=.48\linewidth]{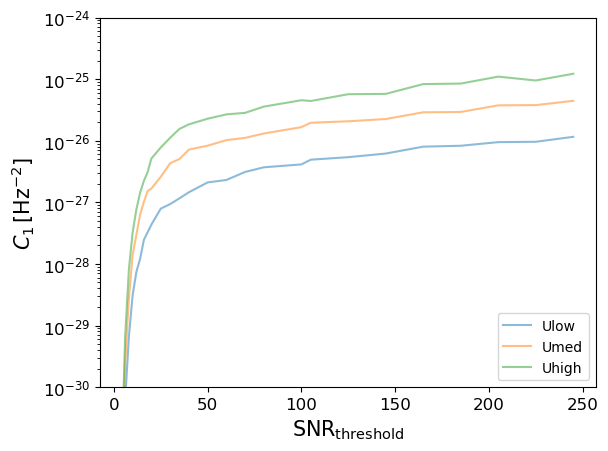}
    \caption{Same as Fig.~\ref{fig:Cl_mass} but for local rates. We apply the uniform mass distribution between $1\,M_{\odot}$ and $2.5\,M_{\odot}$ while we compare different local rates of $80$, $320$ and $810\,{\rm Gpc}^{-3}\,{\rm yr}^{-1}$. Although we calculated 100 realizations and compute angular power up to $l=6$, here we only show the median value of the $l=1$ case out of 100 realizations without error bars for simplicity.}
    \label{fig:Cl_rate}
\end{figure*}

In Fig.~\ref{fig:Cl_rate}, $C_l/C_0$ and $C_l$ are plotted for the models with the same mass distribution but different local rates.
When comparing the local rates of $80$, $320$ and $810\,{\rm Gpc}^{-3}\,{\rm yr}^{-1}$, the population model with a higher local rate gives a larger amplitude of $C_l$. If we have $N$ point sources of the same amplitude, we obtain $C_0\propto\Omegagw^2\propto N^2$ and $C_l \propto N$, then $C_l/C_0\propto1/N$. Therefore, with a higher merger rate, even though the GWB signal becomes more isotropic, the angular power spectrum $C_l$ becomes larger. 

\begin{table}[]
    \centering
    \begin{tabular}{c|c|c}
$l$ & $C_l/C_0 (\times 10^{-6})$ & ${\rm SNR}_{\rm thres}$ \\ \hline
1 & 1.685 & 25 \\
2 & 2.195 & 20 \\
3 & 1.888 & 16 \\
4 & 1.903 & 16 \\
5 & 2.018 & 18 \\
6 & 1.957 & 20
    \end{tabular}
    \caption{The SNR threshold at the minimum $C_l/C_0$ value for each $l$, which is the median value computed from 100 realizations.}
    \label{tab:ClminSNR}
\end{table}

\section{Likelihood estimator and parameter inference}
\label{sec:likelihood_estimator}

In this section, we formulate the likelihood function for the joint estimation of an anisotropic CBC foreground and an isotropic cosmological GWB. Then we derive the Fisher information matrix for the evaluation of the parameter estimation errors of the signals.

\subsection{Likelihood estimator}

The cross-correlation statistic, generally used in previous LIGO, Virgo, and KAGRA anisotropic gravitational wave background searches \cite{O1Anisotropic,O2Anisotropic,O3Anisotropic}, between two detectors $(I,J)$ at time $t$ and frequency $f$, is defined as \cite{Romano_Cornish,Thrane_2009}
\begin{align}
    C_{IJ}(t,f) \equiv \frac{2}{\tau}\tilde{d}_I(t,f)\tilde{d}_J^*(t,f)\;,
\end{align}
where $\tilde{d}_I(t,f)$ is the short Fourier transform of a time-domain detector output $d_I(t)$ with the interval $[t-\frac{\tau}{2},t+\frac{\tau}{2}]$. For simplicity of the notation, we do not write the indice of time segments and frequency bins explicitly. The expectation value of the cross-correlation  \cite{Thrane_2009,Romano_Cornish} is
\begin{align}
    \av{C_{IJ}(t,f)} = \sum_{l,m}\gamma_{IJ,lm}(t;f)\mathcal{P}_{lm}\bar{H}(f)
    \;,\label{eq:Cij_expectation}
\end{align}
where $\mathcal{P}_{lm}$ is the coefficients of the $\mathcal{P}(\bm{\hat{n}})$ in the expansion with spherical harmonics and $\gamma_{IJ,lm}(t;f)$ is the overlap function of the detector network $(I,J)$ for a spherical harmonic mode of $l$ and $m$ \cite{Allen_Ottewill_1997}.

In this analysis, we assume that the correlation signal is the sum of an anisotropic astrophysical foreground and an isotropic cosmological GWB\footnote{In this paper, we do not consider the anisotropies of cosmological backgrounds, since their amplitude is expected to be subdominant compared to the shot noise~\cite{ValbusaDallArmi:2020ifo,Schulze:2023ich}, although cross-correlation with other probes, such as the CMB~\cite{Ricciardone:2021kel}, or non-adiabatic initial conditions~\cite{ValbusaDallArmi:2023nqn,ValbusaDallArmi:2024hwm} could improve the detectability of these primordial signals.}. The expectation value is obtained as the multiple-component version of \eqref{eq:Cij_expectation}:
\begin{multline}
    \label{eq:expectation_multi}
    \av{C_{IJ}(t,f;\bm{\theta})} = \sum_{l=0}^{l_{\rm max}}\sum_{m=-l}^l\gamma_{IJ,lm}(t,f)\mathcal{P}^{({\rm a})}_{lm}\left(\frac{f}{f_{\rm ref}}\right)^{2/3-3}\\
    + \gamma_{IJ,00}(t,f)\mathcal{P}^{({\rm c})}_{00}\left(\frac{f}{f_{\rm ref}}\right)^{\alpha-3} \;,
\end{multline}
where we denoted the set of these parameters, $\mathcal{P}^{({\rm a})}_{lm},\mathcal{P}^{({\rm c})}_{00}$ and $\alpha$, by the vector $\bm{\theta}$. The indices $({\rm a})$ and $({\rm c})$ represent the astrophysical origin and the cosmological origin, respectively. We assume that the cosmological GWB has a power-law spectral dependence on the spectrum index $\alpha$. 

We currently treat $P_{00}^{({\rm a})}$ and $P_{lm}^{({\rm a})}\,(l>1)$ as independent estimators in Eq.~\eqref{eq:expectation_multi}. However, ideally, given the astrophysical population and the sky distribution of the signals, we can theoretically connect $P_{00}^{({\rm a})}$ and $P_{lm}^{({\rm a})}\,(l>1)$ and formulate the relation as functions of the parameters of the model of the astrophysical population. However, it is beyond the scope of this paper and we simply assume that $P_{00}^{({\rm a})}$ and $P_{lm}^{({\rm a})}\quad(l>1)$ are independent at the first stage of the feasibility study of the foreground subtraction method proposed in this paper.

Since the $C_{IJ}(t,f)$ obeys the Wishart distribution, the likelihood function for $C_{IJ}(t,f)$ is written as 
\begin{multline}
    \label{eq:likelihood_shot}
    \ln p(C_{IJ}(t,f)|\mathcal{P}^{({\rm a})}_{lm},\mathcal{P}^{({\rm c})}_{00},\alpha) \\
    \propto -\frac{1}{2}\frac{\left|C_{IJ}(t,f) - \av{C_{IJ}(t,f;\bm{\theta})} \right|^2}{P_{I}(t,f)P_{J}(t,f)}.
\end{multline}

\subsection{Fisher matrix}
\label{sec:Fisher-matrix}
We use the Fisher information matrix to evaluate the estimation error of each parameter. The Fisher matrix is, by definition, derived from
\begin{align}
    \Gamma_{ij} = - \sum_{(I,J)} \sum_{t,f} \left.\av{\frac{\partial^2\ln p({C_{IJ}(t,f; \bm{\theta})})}{\partial\bm{\theta}^*_{i}\partial\bm{\theta}_{j}}}\right|_{\bm{\theta}=\bm{\theta}_{\rm fid}}\;,
    \label{eq:Fisher_matrix}
\end{align}
where $i$ and $j$ are one of the components of the vector 
\beq
\bm{\theta} = \{ \mathcal{P}^{({\rm a})}_{lm},\mathcal{P}^{({\rm c})}_{00}, \alpha \} \;,
\eeq
and $\bm{\theta}_{\rm fid}$ is a set of fiducial parameters. In computing Fisher matrix, we integrate over time and frequency. The integrating time is one year and the frequency range is from 10 to 200 \,Hz. The summation for $(I,J)$ is taken over all detector pairs among ET at the Virgo site, CE with 20 km arms at the LIGO Hanford site, and
CE with 40 km at the LIGO Livingston site. The inverse of the Fisher matrix represents the covariance matrix of the parameters. One can obtain the statistical error of parameter $\bm{\theta}_i$ from the inverse Fisher matrix:
\begin{align}
    \delta\bm{\theta}_{i} = \sqrt{(\Gamma^{-1})_{ii}}.
\end{align}
The correlation coefficients are defined by
\begin{align}
    \label{eq:Fisher_inv_norm}
    c_{ij}\equiv \frac{(\Gamma^{-1})_{ij}}{\sqrt{(\Gamma^{-1})_{ii}(\Gamma^{-1})_{jj}}} \;.
\end{align}
From the fiducial values of $\hat{\mathcal{P}}_{lm}$ and the inverse Fisher matrix, one can derive the unbiased estimator of the angular power spectrum $C_l$~\cite{Thrane_2009},
\begin{align}
    \hat{C}_l = \frac{1}{2l+1}\sum_{m=-l}^l \left[ |\hat{\mathcal{P}}_{lm}|^2-(\Gamma^{-1})_{lm,lm} \right] \;,
    \label{eq:estimator_Cl}
\end{align}
and its covariance
\begin{align}
    \mathrm{Var} \left[\hat{C}_l \right] = \frac{2}{(2l+1)^2}\sum_{m,m'}|(\Gamma^{-1})_{lm,lm'}|^2 \;.
    \label{eq:cov_Cl}
\end{align}

\begin{figure}
    \centering
\includegraphics[width=\linewidth]{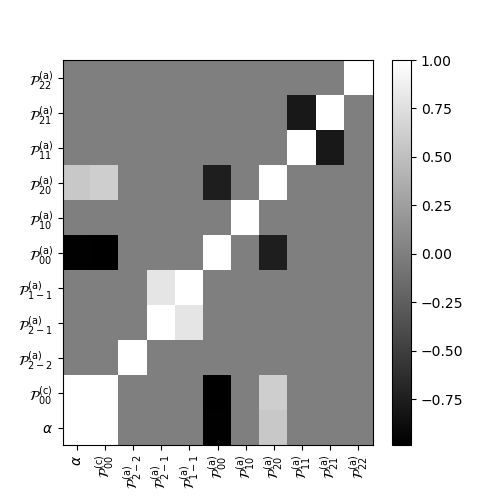}
    \caption{Correlation coefficients between model parameters for the case of $l_{\rm max}=2$.}
    \label{fig:Fisher-inv}
\end{figure}

As a demonstration, here we choose the fiducial values for the Fisher matrix, $l_{\rm max}=2$, $\alpha=0$, and ${\Omega}_{\rm gw}^{({\rm c})}=10^{-15}$, and compute the correlation coefficients and the parameter estimation errors. \figref{fig:Fisher-inv} visualizes the correlation coefficients between the parameters for the case of $l_{\rm max}=2$. $\mathcal{P}^{({\rm a})}_{lm}$ are correlated each other, and more importantly, there is a clear negative correlation between $\mathcal{P}_{00}^{({\rm a})}$ and $\mathcal{P}_{00}^{({\rm c})}$. As described later in \secref{sec:new_estimator}, the correlation between $\mathcal{P}_{00}^{({\rm a})}$ and $\mathcal{P}_{00}^{({\rm c})}$ is crucial when we improve the estimation of $\mathcal{P}_{00}^{({\rm c})}$ by adding the information from $\mathcal{P}_{00}^{({\rm a})}$. 

For the comparison of the results in the later section with and without anisotropic information, we also calculate the parameter estimation errors of $\hat{C}_l$ for the case of $l_{\rm max}=4$ and free $\alpha$. They are computed from Eq.~\eqref{eq:cov_Cl}, and the error of $C_0$ is converted to $\delta \Omega_{\rm GW}$ by
Eq.~\eqref{eq:OmegaGW-C0}. We obtain $\delta{\Omega}_{\rm GW}^{({\rm c})}\sim1.277\times10^{-11}$ and $\delta{\Omega}_{\rm GW}^{({\rm a})}\sim1.326\times10^{-11}$ when we treat $P_{00}^{({\rm a})}$ and $P_{lm}^{({\rm a})}\,(l>1)$ as independent parameters. In the next subsection, we treat $P_{00}^{({\rm a})}$ and $P_{lm}^{({\rm a})}\,(l>1)$ as dependent parameters and compute the parameter estimation errors.

\subsection{Estimation of $\Omega_{\rm GW}^{({\rm a})}$ with the help of anisotropic information}
\label{sec:new_estimator}

In this section, we explain the computation steps to obtain the estimate of an isotropic component, $\Omega_{\rm GW}^{({\rm a})}$, using the anisotropy observation. We consider the estimators for the anisotropy of the astrophysical foreground, $\hat{\mathcal{P}}^{({\rm a})}_{lm}$, the isotropic cosmological background, $\hat{\mathcal{P}}^{({\rm c})}_{00}$, and the tilt of the cosmological background, $\hat{\alpha}$. The errors associated to these estimators have been obtained by inverting the Fisher matrix introduced in Eq.~\eqref{eq:Fisher_matrix}, which are denoted by $\delta \hat{\mathcal{P}}^{({\rm a})}_{lm}$, $\delta \hat{\mathcal{P}}_{00}^{({\rm c})}$, and $\delta\hat{\alpha}$. For each $l$-mode, we evaluate the angular power spectrum of the astrophysical foreground with the estimators $\hat{C}_l$ defined in Eq.~\eqref{eq:estimator_Cl} and its error $\delta\hat{C}_l$ by taking the square root of Eq.~\eqref{eq:cov_Cl}. Since $\mathcal{P}_{00}^{({\rm a})}$ and $\mathcal{P}_{00}^{({\rm c})}$ are strongly correlated, adding the anisotropic information of a source distribution is essential to break the degeneracy. Particularly, in the case of $\alpha=2/3$, there is a complete degeneracy between the spectral index of the cosmological background and the astrophysical foreground in the low-frequency spectrum. The anisotropic information can break the degeneracy in principle.

Given the mass and redshift distributions of the astrophysical population, one can calculate the ratio of the power in the monopole mode to that in the $l$-mode $(l \geq 1)$,
\begin{equation}
    A_{\rm th} \equiv \frac{C_{0,\rm th}}{C_{l, \rm th}} \,,
\label{eq:monopole_anisotropies_AGWB}
\end{equation}
and use this ratio as a conversion factor from the higher-multipole components to the monopole component. Note that $A_{\rm th}$ is independent of $l$ and does not have the subscript $l$. Then the monopole amplitude of the astrophysical foreground is estimated independently from each estimate of the angular power spectrum $\hat{C}_l$ by
\begin{align}
    \hat{\Omega}_{{\rm GW},(l,{\rm new})}^{({\rm a})} = \frac{2\pi^2}{3H_0^2} {f_{\rm ref}^3} \sqrt{4\pi A_{\rm th}\,\hat{C}_l} \,. 
    \label{eq:estimator-OmegaGW-astro} 
\end{align}

In general, the parameter $A_{\rm th}$ is a function of astrophysical population model parameters, and their observational errors should be taken into account. However, in this paper, as a first step of the demonstration of the method, we fix the astrophysical population for simplicity and adopt the median value of $A_{\rm th}$ calculated from 100 realizations of the source distribution on the sky\footnote{In reality, one realization in our Universe includes the cosmic variance. However, it can be reduced by subtracting louder signals individually. The remaining signals have less variance in amplitude because of a larger number of signals with modest amplitude. The partial subtraction also affect the inference of the astrophysical population model parameters. For a complete analysis including the statistical variance from the astrophysical population, a more complicated construction of the likelihood function is necessary.}. The error associated to the estimator, $\hat{\Omega}_{{\rm GW},(l,{\rm new})}^{({\rm a})}$, is obtained by the propagation of errors,
\begin{equation}
    \label{eq:delta_OmegaGW}
    \delta\hat{\Omega}_{{\rm GW},(l,{\rm new})}^{({\rm a})} = \frac{2\pi^2}{3H_0^2} {f_{\rm ref}^3} \sqrt{\frac{\pi A_{\rm th}}{\hat{C}_l}} \delta \hat{C}_l\,.
\end{equation}

Once we have a set of the independent $l$-mode amplitude estimators of the astrophysical foreground, it is possible to combine them into an optimal estimator, which is expected to have a lower variance. The optimal estimator is obtained by taking the weighted average of the individual estimators and is defined by
\begin{align}
    \hat{\Omega}_{\rm GW,new}^{(\rm a)} = \frac{\sum_l\left\{\delta\hat{\Omega}_{{\rm GW},(l,{\rm new})}^{({\rm a})}\right\}^{-2}\hat{\Omega}_{{\rm GW},(l,{\rm new})}^{({\rm a})}}{\sum_l\left\{\delta\hat{\Omega}_{{\rm GW},(l,{\rm new})}^{({\rm a})}\right\}^{-2}} \,,
    \label{eq:optimal-estimator}
\end{align}
where the sum is taken from $l=0$. The maximal $l$ in the inference is determined by the angular resolution of the detector network. For our case of three 3G detectors, $l_{\rm max}=4$ is a reasonable choice, because the detector network is sensitive roughly up to $l_{\rm max}=4$ and the Fisher matrix starts being ill-conditioned above $l_{\rm max}=4$, which increases the errors of all $C_l$.

\begin{figure}
    \centering
    \includegraphics[width=\linewidth]{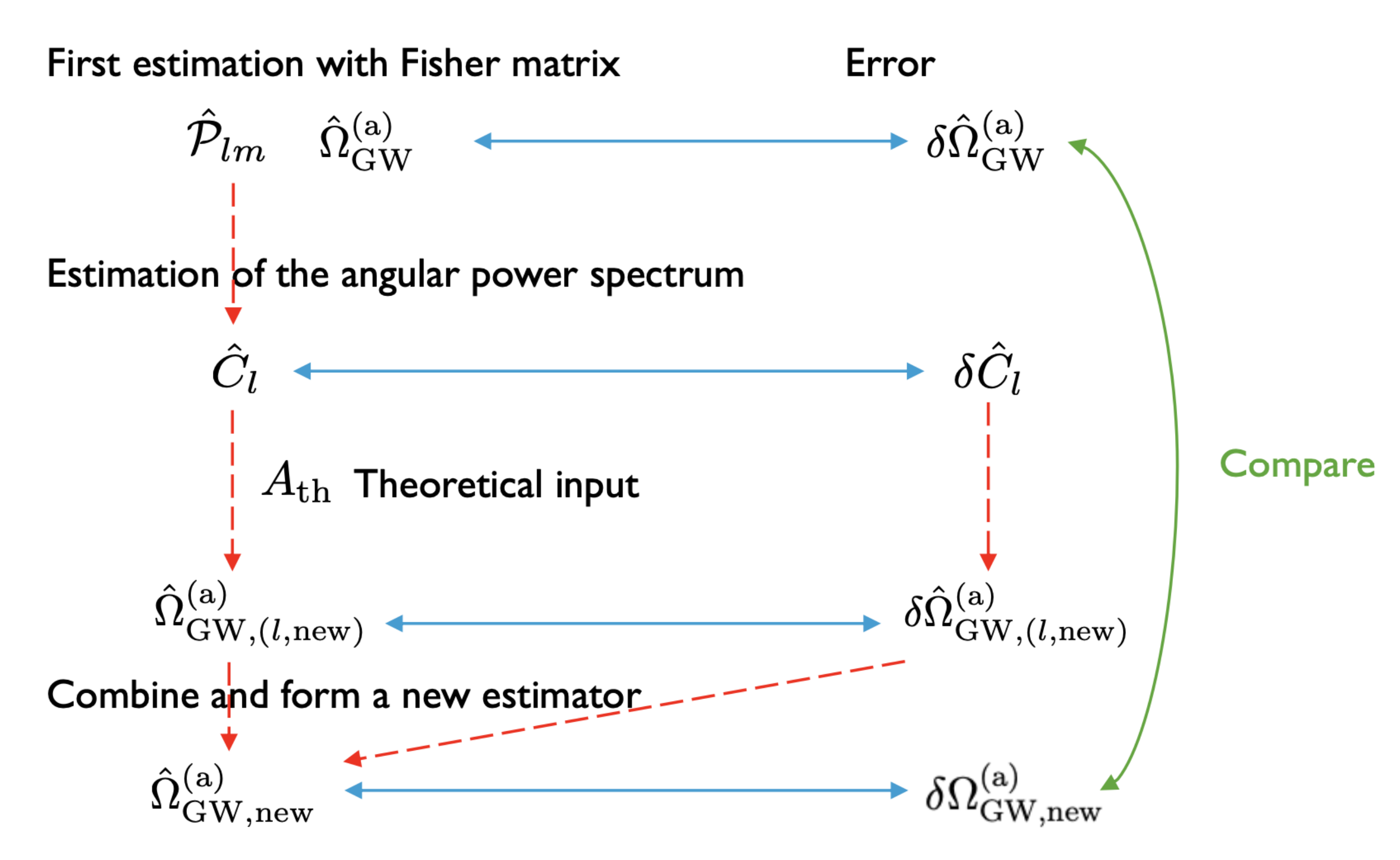}
    \caption{Computation steps to construct $\hat{\Omega}_{\rm GW,new}^{(\rm a)}$.}
    \label{fig:new-estimator}
\end{figure}

The computation steps are visualized in \figref{fig:new-estimator}. With improvement in the estimation of ${\Omega}_{\rm GW}^{(\rm a)}$, we can also improve the estimation of ${\Omega}_{\rm GW}^{(\rm c)}$ because they are strongly correlated, as shown in \figref{fig:Fisher-inv}. We define the {\it improvement fraction} to check how much the estimation error is improved by adding the information from the observation of anisotropy as
\begin{align}
    \label{eq:improvement_fraction}
    {\rm improvement\ fraction} =
    \frac{\delta\hat{\Omega}_{\rm GW}^{(\rm a)}-\delta\hat{\Omega}_{\rm GW,new}^{(\rm a)}}{\delta\hat{\Omega}_{\rm GW}^{(\rm a)}} .
\end{align}

\section{Results}
\label{sec:results}

\begin{figure}
    \centering
    \includegraphics[width=\linewidth]{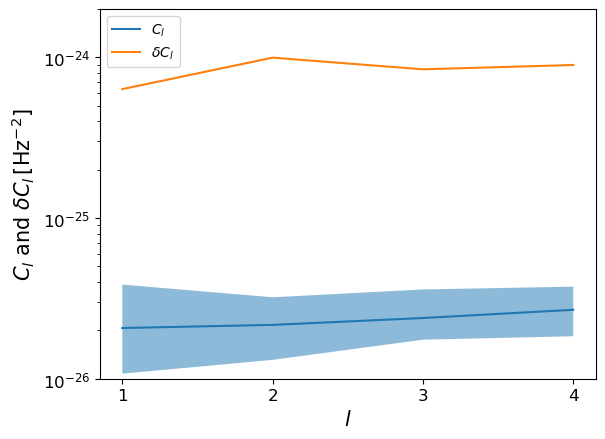}
    \caption{$C_l$ values up to $l=4$ from BNS background with the astrophysical population of pop Umed model (blue). The shaded blue region is cosmic variance simulated from 100 realization taking 25th and 75th percentiles as lower and upper bounds. $\delta C_l$ is the instrumental error from the Fisher matrix calculation by taking $l_{\rm max}=4$ and assuming the detector network of three third-generation detectors as described in \secref{sec:SNR computation}.}
    \label{fig:angular-power-detector-error}
\end{figure}

We choose the pop Umed model as a fiducial scenario and compute the improvement fraction which is defined by Eq.~\eqref{eq:improvement_fraction}. We also use the median value from 100 realizations of the sky distributions of sources in Eq.~\eqref{eq:delta_OmegaGW} to calculate $\delta\Omega^{({\rm a})}_{{\rm GW},(l,{\rm new})}$. In reality, the distribution of $C_l$ values differs, depending on the sky distributions of BNS in the observation period. Therefore, the improvement fraction could change better or worse depending on a realization of the $C_l$ values during the observation. In Fig.~\ref{fig:angular-power-detector-error}, we show the values of $C_l$ and the instrumental error $\delta C_l$, assuming the detector network of three third-generation detectors as described in \secref{sec:SNR computation}. We have found that our three-detector network has sensitivity to $C_l$ down to $10^{-25}\sim 10^{-24}\,\,{\rm Hz}^{-2}$, with the highest sensitivity to the dipole anisotropy. However, the typical amplitude of $C_l$ from the BNS foreground is $\sim 10^{-26}\,\,{\rm Hz}^{-2}$ in the fiducial scenario {and cannot be measured well.}

From Eqs.~\eqref{eq:estimator-OmegaGW-astro}--\eqref{eq:optimal-estimator}, we compute the estimate of $\delta\hat{\Omega}_{\rm GW, new}^{(\rm a)}$, including the information from the anisotropic source distribution on the sky. Comparing with the error computed in Sec.~\ref{sec:Fisher-matrix}, we compute the improvement fraction defined in Eq.~\eqref{eq:improvement_fraction} as a function of SNR threshold for signal subtraction and plot it in blue in Fig.~\ref{fig:imp_rate_factor}. The estimation of the monopole amplitude of the astrophysical foreground is hardly improved by adding the information of the anisotropy, at most $\sim 10^{-5}$ even if almost all louder signals are kept. This is because the measurement errors of $C_l$ are rather large for ground-based detectors such as ET and CE, and the information from the anisotropy is very limited.

\section{Discussions}
\label{sec:Discussion}

\begin{figure}
   \centering
\includegraphics[width=\linewidth]{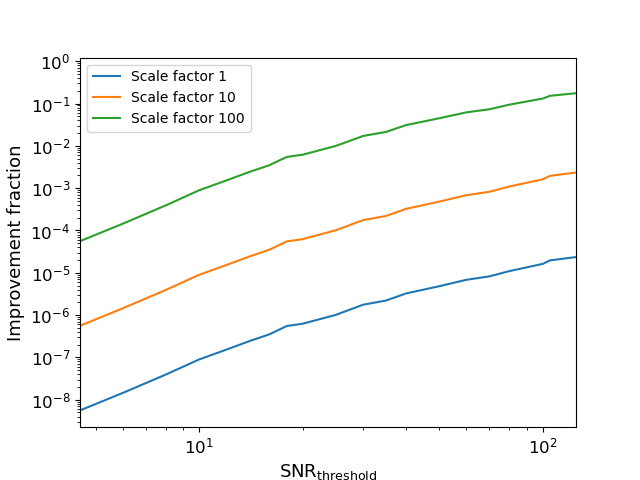}
    \caption{The figure of improvement fraction with the choice of SNR threshold. To demonstrate how much improvement factor get affected by the better detector sensitivity, we multiply the sensitivity of three detector network by the scale factors given in the graph. If we lower the SNR threshold the remaining BNS signals creates different (tends to be lower) angular power. Therefore, the relative size of estimation error towards the value of angular power varies with the threshold.}
    \label{fig:imp_rate_factor}
\end{figure}

As explained in \secref{sec:results}, our detector network is expected to detect no anisotropy of the astrophysical foreground from shot noise after perfectly subtracting louder signals. Therefore, to discuss the effectiveness of our method, we artificially scale the sensitivity to the angular power spectrum, $\delta C_l$ for $l>1$, by up to a factor of $100$ and see how much the estimation error of ${\Omega}_{\rm GW}^{(\rm c)}$ is improved. We should note that we did not scale the sensitivity to the isotropic component at all. The result is shown in \figref{fig:imp_rate_factor}. The SNR threshold on the horizontal axis means that signals with SNR above the threshold are perfectly subtracted. We further set the maximum SNR threshold to $125$ above which we have less than 500 signals and the standard assumptions of the stochastic background, Gaussianity and stationary, are broken due to the small number of very loud signals. The angular power spectrum $C_l$ decreases with lowering the SNR threshold. When the anisotropic signal is weaker, we have less additional information about the amplitude of the isotropic signal from the anisotropy and the improvement factor becomes worse. Since the scale factor just multiplies the sensitivity of the detector network towards the angular power, the improvement fraction linearly scales with the scale factor. We found that $\Omega^{(\rm a)}_{\rm GW}$ is estimated with the error smaller by about $17.5\%$ with the scaling of sensitivity by a factor of $100$ by adding the information from the anisotropy of the astrophysical foreground up to $l=4$.

In this work, we have neglected the impact of the kinetic dipole induced by our peculiar motion with respect to the rest frame of the astrophysical sources~\cite{Cusin:2022cbb, ValbusaDallArmi:2022htu} as well as the contribution from circular polarization, which can exist only for an anisotropic source distribution~\cite{ValbusaDallArmi:2023ydl}. These effects could be observed by future interferometers and may improve our conclusions, but we leave the analysis for future work.

\section{Conclusions}
In this paper, we propose a method to estimate the isotropic component of a cosmological background with the use of the information from the anisotropic distribution of the astrophysical CBC background. We first formulate the likelihood function for the mixture of the cosmological and astrophysical backgrounds in \eqref{eq:likelihood_shot}. Then using the Fisher matrix, we evaluate the estimation errors of the isotropic and anisotropic components and show how much we could improve the estimation error of the isotropic component by utilizing the information on an anisotropic astrophysical background.

Unfortunately, we found that the anisotropy from the BNS background after subtracting louder signals is too faint to observe with the 3G generation detector network and our method cannot improve the estimation error of a cosmological background. However, in order to quantify the effect of our method, we artificially increase the sensitivity to the angular power spectrum and see the reduction of the estimation error, showing a better estimate by about $17.5\%$ with 100 times better sensitivity for the case of SNR threshold of $125$. The estimation error of the foreground becomes the residual energy density after we subtract the estimated foreground energy density from the observed energy density. This residual energy density will limit the detection of other stochastic GWB. It is still possible that we can use our method with a space-based detector such as DECIGO. Capurri {\it et al.}~\cite{Capurri_2023} showed that the shot noise in the angular power spectrum of BNS signals is louder than the detector noise power spectrum of DECIGO by more than two orders of magnitude. We would expect that the anisotropic information of the astrophysical foreground plays a crucial role to estimate the isotropic cosmological background. For such a case, we need to construct the joint likelihood function for individual CBC signals and their unresolved component, as proposed by Zhong {\it et al.}~\cite{Zhong:2025qno}. Therefore, the application of our method to space-based detectors is more complicated and we leave it as a future work.\\

\begin{acknowledgments}
This article is supported by the JSPS Grants-in-Aid for the JSPS Research Fellow (KAKENHI) grant number JP22KJ1033 and the Grant-in-Aid for Scientific Research(A) grant number JP18H03698. A.~N. is supported by JSPS KAKENHI Grant Nos. JP23K03408, JP23H00110, and JP23H04893.
The authors are grateful for computational resources provided by the LIGO Laboratory and supported by National Science Foundation Grants PHY-0757058 and PHY-0823459. We thank also Kipp Cannon (RESCEU, Tokyo) for valuable discussions on this work.
\end{acknowledgments}

\bibliography{reference}
\end{document}